# Room-Temperature Superconductivity at 298 K in Ternary La-Sc-H System at High-pressure Conditions

Yinggang Song[1], Chuanheng Ma[1], Hongbo Wang[2], Mi Zhou[1], Yanpeng Qi[3], Weizheng Cao[3], Shourui Li[4], Hanyu Liu[1,2,5], Guangtao Liu[1,*], Yanming Ma[6,1,2,*]

[1] *Key Laboratory of Material Simulation Methods and Software of Ministry of Education, College of Physics, Jilin University, Changchun 130012, China*

[2] *State Key Laboratory of Superhard Materials College of Physics, Jilin University, Changchun 130012, China*

[3] *State Key Laboratory of Quantum Functional Materials, ShanghaiTech Laboratory for Topological Physics, School of Physical Science and Technology, ShanghaiTech University, Shanghai 201210, China*

[4] *Institute of Fluid Physics, China Academy of Engineering Physics, Mianyang 621900, China*

[5] *International Center of Future Science, Jilin University, Changchun 130012, China*

[6] *College of Physics, Zhejiang University, Hangzhou 310027, China*

**Room-temperature superconductor has been a century-long dream of humankind. Recent research on hydrogen-based superconductors (e.g., $CaH_6$, $LaH_{10}$, etc.) at high-pressure conditions lifts the record of superconducting critical temperature ($T_c$) up to ~250 kelvin. We here report the experimental synthesis of the first-ever room-temperature superconductor by compression on a mixture of La-Sc alloy and ammonia borane at pressures of 250-260 gigapascals (GPa) via a diamond anvil cell by a laser-heating technique. Superconductivity with an onset temperature of 271-298 kelvin at 195-266 GPa is observed by the measurement of zero electrical resistance and the suppression of $T_c$ under applied magnetic fields. Synchrotron X-ray diffraction data unambiguously reveal that this superconductor crystallizes in a hexagonal structure with a stoichiometry $LaSc_2H_{24}$, in excellent agreement with our previous prediction[1]. Through thirteen reproducible experimental runs, we provide solid evidence of the realization of a room-temperature superconductor for the first time, marking a milestone in the field of superconductivity.**

**Introduction**

Since the first discovery of superconductivity in mercury, scientists have devoted persistent efforts to discovering superconductors with higher superconducting critical temperatures ($T_c$). In 1985, Nobel laureate V. L. Ginzburg emphasized "high-temperature superconductivity" as one of the most important and intriguing problems in macrophysics[2]. Over the past decades, major advances have been achieved with the development of unconventional superconductors, including cuprates[3-5], iron-based materials[6,7], nickelate superconductors[8-10], and especially conventional hydrogen-based superconductors that have pushed $T_c$ to unprecedented levels (as high as 250 K)[11-13]. Nevertheless, the realization of room-temperature superconductivity remains a central goal and a major challenge in modern science.

Although highly compressed hydrogen has long been proposed as a high-temperature superconductor owing to its lightest atomic mass and strong electron-phonon interactions[14], direct experimental evidence for its metallic phase remains elusive to date[15,16]. As alternatives, hydrogen-based compounds have emerged as promising candidates for high-temperature superconductivity at relatively lower pressures[17,18]. Progress in this field was slow for decades, largely because the specific hydrides capable of high-$T_c$ superconductivity had not yet been discovered, until the advent of crystal structure prediction methods transformed the landscape. The "clathrate hydride" paradigm was originally proposed for $CaH_6$ in 2012 (ref.[19]) and experimentally verified a decade later, with superconductivity above 200 K (refs [20,21]). In this structure, hydrogen transforms from bonding to antibonding states, resulting in a high density of states at the Fermi level and enhancement of electron-phonon coupling. Building on this insight, a class of high-temperature superconducting rare-earth clathrate hydrides were discovered, including $LaH_{10}$ (refs [11,12,22-25]), $YH_9$ (refs [22,26,27]), $YH_6$ (refs [22,26-29]), and $CeH_{9-10}$ (ref.[30]), with $LaH_{10}$ setting the $T_c$ record (250 K at ~180 GPa)[11,12]. Other binary hydrides, such as covalent H-S compounds with $T_c$ up to 203 K (refs [13,31-34]), have also been extensively explored[35].

Compared with binary hydrides, ternary hydrides have attracted growing interest owing to their greater elemental diversity and structural complexity[36-38], which may open new avenues toward room-temperature superconductivity[36,39]. For instance, metastable $Li_2MgH_{16}$ has been predicted to exhibit "hot" superconductivity with $T_c$ of 351-473 K under multimegabar pressure[40]. In addition, several other high-$T_c$ hydrides, including $MB_xH_y$ (refs [41-43]), $MBeH_8$ (refs. [44,45]), and $Mg_2IrH_{6-7}$ (refs [46-48]), have been calculated to be stable under comparatively lower pressures. Experimentally, progresses have been achieved by incorporating a third element into the La-H system, giving rise to two types of superconducting ternary hydrides. The first involves doping binary hydride to form non-stoichiometric alloy superhydrides (e.g. $(La,Y)H_6$, $(La,Y)H_{10}$ (ref. [49]), $(La,Ce)H_{9-10}$ (refs [50,51]), $(La,Nd)H_{10}$ (ref. [52]), and $(La,Ca)H_{10}$ (ref. [53])), where dopant atoms randomly substitute the metal positions or occupy interstitial sites within the parent binary hydrides. This strategy, however, introduces no new structure and has yielded only limited improvements in superconductivity[51-53]. The second pathway involves the formation of entirely new ternary structural prototypes, such as recently synthesized $LaBeH_8$ and $LaB_2H_8$ (refs [54,55]), featuring a La-based framework integrated with $XH_n$ units within the La sublattice. Although their $T_c$s do not surpass those of certain binary hydrides, these compounds demonstrate the feasibility of designing and synthesizing ternary hydrides with diverse structural motifs capable of supporting high-temperature superconductivity.

Very recently, our theoretical study employing the crystal structure prediction (CALYPSO) method proposed a ternary high-$T_c$ superhydride, $LaSc_2H_{24}$ (ref. [1]), featuring hexagonal $P6/mmm$ symmetry. The introduction of Sc, which shares a similar valence electron configuration as La but possesses a lighter mass and a smaller atomic radius, resulted in the formation of two previously unreported hydrogen cages: $H_{24}$ surrounding Sc and $H_{30}$ surrounding La. These two novel hydrogen frameworks were predicted to produce an exceptionally large hydrogen-derived density of states at the Fermi level, giving rise to an exceptionally high $T_c$ of up to 316

K at high pressure. Crucially, LaSc$_2$H$_{24}$ is thermodynamically stable across experimentally accessible pressures, bringing the prospect of room-temperature superconductivity closer than ever before.

Inspired by the above prediction[1], we synthesized a ternary clathrate La-Sc alloy hydride and investigated its superconducting properties under high pressure. *In-situ* X-ray diffraction (XRD) experiments revealed that the compound crystallizes in a hexagonal structure, in excellent agreement with the predicted *P6/mmm* LaSc$_2$H$_{24}$ structure, and remains stable down to 194 GPa. Superconductivity, with a maximum onset temperature ($T_c^{onset}$) of 298 K, was confirmed by both the observation of zero electrical resistance and the suppression of $T_c$ under applied magnetic fields. These results announce the realization of a room-temperature superconductor and may herald the emergence of superconductors with even higher $T_c$.

**Results**

**Synthesis of the ternary La-Sc-H compound**

Based on the theoretically predicted stoichiometry, a homogeneous La-Sc alloy with an elemental ratio of ~1:2 (Extended Data Table 1) and NH$_3$BH$_3$, a common hydrogen source in high-pressure hydride synthesis [11,20,26,30,50,54], were used as precursors. High pressure was generated using a diamond anvil cell (DAC), and laser heating provided the thermal energy necessary to overcome the reaction barrier. After irradiation, the samples turned black (Fig. 1 (b) and Extended Data Fig. 1) and exhibited a pressure drop, suggesting that the intended chemical reaction had occurred, accompanied by a probable volume collapse.

Since the target compound LaSc$_2$H$_{24}$ has been calculated to be thermodynamically stable at pressure as low as 167 GPa (ref. [1]), initial synthesis attempts were carried out below 200 GPa. The samples were pressurized at room temperature and subsequently heated for one minute. Although some products were non-superconducting, superconducting transitions with onset temperatures around 250 K (Extended Data Fig. 2) were observed in three experimental runs at 183-195 GPa. These were attributed either to binary LaH$_{10}$ (refs [11,12,56]) or an unidentified ternary La-Sc-H phase, neither of which showed a significant enhancement in $T_c$, suggesting that the target $LaSc_2H_{24}$ had not yet formed. Approximately ten further attempts to synthesize $LaSc_2H_{24}$ between 200 and 250 GPa also did not yield higher $T_c$. Consequently, the synthesis pressure was increased to ~260 GPa, where formation of $LaSc_2H_{24}$ is more favorable, as calculations indicate improved energetic stability relative to decomposition products[1]. At this pressure, we observed a pronounced enhancement of superconductivity, with $T_c$ significantly exceeding the ~250 K reported in $LaH_{10}$, although whether the La-Sc-H compound with high $T_c$ can be synthesized at lower pressure remains an open question.

**Superconducting temperature of the La-Sc-H compound**

We first reproduced the superconductivity in binary La-H system at 170 GPa, observing a $T_c$ of 248 K with a clear transition to zero resistance (Extended Data Fig. 2), in excellent agreement with previous report on $LaH_{10}$ (ref. [12]). This validates the reliability of our cryogenic system and electrical transport measurement. We then probe the superconductivity of the synthesized La–Sc–H compound near 250 GPa. Representative resistance-temperature curves from independent runs in cells 1, 3, 4, and 5 are shown in Figure 2(a). Abrupt resistance drops were observed at 295 K (245 GPa), 283 K (253 GPa), 298 K (260 GPa), and 295 K (262 GPa) in cells 1, 3, 4, and 5, respectively, signaling the onset of the superconducting transitions. Notably, the "zero resistance", characterized by oscillations around zero, was detected in cells 4 and 5 (insert of Fig. 2(a)), thereby ruling out a temperature-induced phase transition and providing one of the compelling evidence of superconductivity. Besides, superconductivity with lower $T_c$s values of 272 K at 256 GPa and 271 K at 260 GPa was observed in cells 4* and 5* (Extended Data Figs. 7 and 8). Furthermore, in all electrical transport experiments, we confirmed that all four electrodes remained conductive, eliminating the possibility of a short producing the "false zero resistance" phenomenon postulated by Hirsch[57].

Given the considerable scientific significance and interest of determining whether pressure can tune or enhance $T_c$, the pressure dependence of $T_c^{onset}$ is investigated as shown in Fig. 2(b). In cells 1 and 3, the $T_c^{onset}$s decrease with increasing pressure, from 295 K (245 GPa) to 289 K (254 GPa), and from 283 K (253 GPa) to 281 K (257 GPa), respectively. In cells 4 and 5, $T_c^{onset}$'s consistently exceeds 290 K and shows no clear pressure dependence within the range of 195-266 GPa. At 195 GPa, the exhibits anomalous behavior, including an abnormal jump during the superconducting transition and an unphysical negative resistance between 230 and 277 K, rather than fluctuating near zero (Extended Data Fig. 5(a)). These features suggest a circuit abnormality, likely caused by partial decomposition of this superconducting phase at this pressure.

It is worth noting that $T_c^{onset}$ varies among different cells at the same pressures. This variation may arise from sample inhomogeneity, such as a slight difference in hydrogen concentration during synthesis and pressure gradients under ultrahigh pressures, as commonly observed in other reported high-$T_c$ hydrides[12,30,54]. Furthermore, recent work on diffusion-driven transient hydrogenation in metal superhydrides under extreme conditions has demonstrated dynamic de-hydrogenation and gradual decomposition over several days, which can also affect the $T_c$[58]. These considerations suggest that, upon decompression, factors beyond pressure may influence the $T_c$ of this hydride. Despite such variations, the observed $T_c$ values represent the highest reported to date and are consistent with our previous theoretical prediction of 296 K at 250 GPa based on anharmonic approximation[1], which significantly impacts dynamic stability, phonon frequencies, and superconductivity[25,33,34].

**Suppression of superconductivity under external magnetic fields**

The Meissner effect, which describes the expulsion of magnetic flux from a superconductor, serves as key evidence for superconductivity alongside zero resistance. Recent advances in magnetic measurement techniques have enabled detection of this effect even within diamond anvil cells[59]. However, in ultra-high-pressure experiments, the minute sample size (~10-20 $\mu$m) produces very weak signals, restricting magnetic measurements. Nevertheless, superconductivity can be verified by the suppression of $T_c$ under applied external magnetic fields, a method commonly employed in high-pressure superconducting hydride studies[12,20,30,54,60]. $T_c$ decreases progressively with increasing fields due to the breaking of Cooper pairs via orbital and spin-paramagnetic effects. As shown in Fig. 3(a), $T_c$ decreased by about 11 K, from 296 K to 285 K under a magnetic field ($\mu_0 H$) of 0-9 T at 219 GPa in cell 4. Importantly, this suppression was consistently reproduced in other different cells 3, 4*, and 5 (Extended Data Figs. 8, 9, and 10), strongly confirming the superconducting nature of the synthesized La-Sc-H compound.

To estimate the upper critical magnetic field ($\mu_0 H_{c2}(T)$), we extrapolated two sets of $T_c$ values, defined as the temperature at which the resistance drops to 90% and 50% of the normal-state value at 219 GPa. The extrapolated $\mu_0 H_{c2}(0)$ ranges are 89-114 T using the Ginzburg-Landau (GL)[61,62] model, 122-156 T using the Werthamer-Helfand-Hohenberg (WHH)[63] model, and 177-228 T using a linear fitting commonly applied in compressed superhydrides[64,65]. The GL coherence length, derived from $H_{c2}(0)$, is 1.70-1.92 nm, indicating that the synthesized superconducting hydride is a typical type-II superconductor.

**Crystal structure of the synthesized La-Sc-H compounds**

To elucidate the crystal structure of the synthetic La-Sc-H products and gain insights into the mechanism of room-temperature superconductivity, we performed *in-situ* XRD measurements in cells 1, 2, and 5. Notably, high-temperature superconducting transitions of 289-295 K were observed in cells 1 and 5 (Extended Data Figs. 3 and 6). As shown in Fig. 4(a), at 254 GPa, four peaks at 8.44°, 13.56°, 14.64°, and 20.01° in cell 1 can be indexed by a hexagonal close-packed (*hcp*) lattice with cell parameters of $a$ = 4.86(4) Å and $c$ = 3.35(6) Å, which correspond well with the theoretical *P6/mmm* structure of stoichiometric $LaSc_2H_{24}$ ($a$ = 4.831 Å and $c$ = 3.341 Å at 250 GPa) (ref. [1]). These peaks correspond to the (1 0 0), (1 0 1), (1 1 0), and (2 0 1) planes, respectively. Two additional weak peaks at 16.92° and 18.11° assigned to the (2 0 0) and (1 1 1) planes are consistent with the calculated XRD pattern (Fig. 4(a)), but were not apparent in the structural refinement. Their presence, however, can be discerned in the raw pattern (Extended Data Fig. 12), further confirming that the synthesized compound possesses *hcp* symmetry. Moreover, the LaSc$_2$H$_{24}$ phase was excellently identified from 13 diffraction peaks in another independent cell 5 (Extended Data Fig. 14), where superconductivity with $T_c$ above 290 K was verified. The experimentally refined cell parameters ($a$ = 4.85(9) Å and $c$ = 3.35(5) Å) and volume are perfectly consistent with the simulation values, indicating the reliability and reproducibility of the high-$T_c$ phase.

To assess the structural stability of this synthesized *hcp* LaSc$_2$H$_{24}$, decompression XRD was carried out in cell 2. As shown in Extended Data Fig. 13, diffraction peaks corresponding to the *P*6/*mmm* structure were identified at 266 GPa and gradually shifted to lower angles as the pressure decreased to 194 GPa. Below 194 GPa, the (1 0 0) and (1 1 0) peaks began to split, and the intensity of (1 0 1) peak gradually decreased, indicating the onset of lattice instability or the partial decomposition, consistent with the abnormal superconducting behavior observed at 195 GPa in electrical measurements. Using the refined crystal structure parameters at each pressure, we constructed the equation of state (EOS) for the synthesized LaSc$_2$H$_{24}$ (Fig. 4(b)). Experimental EOS data from cell 2 follow a continuous curve without noticeable discontinuities, indicating good stability and reproducibility across the pressure range. Notably, the experimental EOS is consistent with, but slightly higher than, the theoretical curve of the *P*6/*mmm* LaSc$_2$H$_{24}$ (Ref. [1]), which may reflect a typical difference between experiment and theory or a slight deviation in the hydrogen content from the ideal stoichiometry, a phenomenon commonly observed in other compressed hydrides[24,51,65]. In addition, a recent experimental work claimed a synthesis of LaSc alloy hydride with an equimolar metal ratio at ~190 GPa, exhibiting a resistance drop at 274 K[66]. This phase was proposed to adopt a substituted-type hexagonal (La,Sc)H$_{6-7}$, with a diffraction pattern distinct from that observed in this work (Extended Data Fig. 15).

**Discussion**

**Hydrogen in high-$T_c$ LaSc$_2$H$_{24}$**

Figure 5 shows the crystal structure of the $P6/mmm$-structured $LaSc_2H_{24}$, where La and Sc atoms occupy lattice sites of an $MgB_2$-type structure and are each surrounded by $H_{30}$ cages and $H_{24}$ cages, both of which are not observed in known clathrate or zeolite networks. While the diffraction peaks can be well refined with the $LaSc_2H_{24}$ structure, they primarily reveal the positions of the metal sublattice. Due to the limited flux of current X-ray sources, resolving the exact positions of light H atoms remains challenging, an inherent limitation of XRD-based structure determination of superhydrides under ultrahigh pressures. Nevertheless, the hydrogen concentration can be estimated from lattice volume expansion caused by hydrogen, a method previously applied to other polyhydrides[12,20,26,30,50,54]. Although the atomic volumes of La, Sc, and H atoms in the ternary compound are not directly known, they can be inferred from their high-pressure elemental phases[67-69]. Taking an example, at 254 GPa, the refined volume ($V_{lattice}$) of synthesized La-Sc-H compound is 68.788 $Å^3$ /f.u., and the estimated atomic volumes $V_{La}$, $V_{Sc}$, and $V_H$ are ~12.990, ~8.196, and ~1.620 $Å^3$ /atom, respectively. Based on these values, the hydrogen contents ($n$) is calculated as $n = (V_{lattice} - V_{La} - 2V_{Sc})/V_H = 24.3$, almost close to the ideal value of 24.

We also observed probable de-hydrogenation of the synthesized $LaSc_2H_{24+x}$ (x~-1.1–1.4) during decompression, a behavior similar to that reported in other superconducting hydrides such as $LaH_{10\pm x}$ (refs [12,24]), where the experimentally estimated hydrogen content often deviates from the ideal stoichiometry and varies across different cells and pressures. As shown in Fig. 4(b), the experimental EOS curve intersects the dashed curve representing the unreacted elemental assemblage at ~250 GPa, suggesting hydrogen loss upon decompression. The estimated hydrogen contents decrease from 25.4 at 266 GPa to 22.9 at 194 GPa, corresponding to x = -1.1 to +1.4 in $LaSc_2H_{24+x}$ and a deduced de-hydrogenation rate of -0.03 atom/GPa (Extended Data Table 3). Nevertheless, direct characterization of hydrogen remains limited, and relevant high-pressure techniques are still developing. Emerging methods such as high-pressure nuclear magnetic resonance[70] may provide further insights into hydrogen in the crystal structure of $LaSc_2H_{24}$ and its role in facilitating high-temperature superconductivity.

**Significant role of Sc in high-$T_c$ $LaSc_2H_{24}$**

Finally, we discuss the underlying mechanism behind room-temperature superconductivity in $LaSc_2H_{24}$. Among rare earth elements, Sc is unique in the La-X-H ternary superhydride family due to its lightest atomic mass, smallest atomic radius, and a valence electronic configuration ($4s^23d^1$) similar to that of La ($6s^25d^1$). A lighter atomic mass is known to favor higher $T_c$ in conventional superconductors. Moreover, the atomic radius ratio of La to Sc is estimated to be about 1.17 at 254 GPa, surpassing the 15% limit specified by the Hume-Rothery rule[71]. In contrast, previous ternary superhydrides such as La-Ce-H[50,51] and La-Y-H[49,60], exhibit the minimal atomic size mismatch (e.g., La/Ce~1.03 (ref. [50])), leading to a disordered solid solution where the two metals randomly occupy equivalent sublattice positions, leaving the hydrogen clathrate framework largely unaltered. In the present work, introduction of Sc into the La-H binary system under high pressure yields a fundamentally different outcome: a nested, composite structure. La and Sc occupy distinct $MgB_2$-type lattice sites, stabilizing two previously unreported $H_{30}$ and $H_{24}$ cages. These unique hydrogen frameworks form a weakly covalent, nonmolecular hydrogen network, directly enhancing the Fermi-level density of states at the and electron-phonon coupling. Moreover, Sc lacks localized $f$ electrons, unlike Ce or Nd, thereby avoiding the magnetic scattering that can suppress $s$-wave superconductivity[50,52]. Collectively, these analyses highlight the crucial role of Sc in stabilizing novel hydrogen clathrates and enabling room-temperature superconductivity in $LaSc_2H_{24}$, with a dramatic $T_c$ enhancement compared to $LaH_{10}$. This insight naturally points to an exciting future direction: doping the La-Sc-H system with other appropriate elements, guided by the advanced crystal structure prediction method, may further enhance $T_c$.

With the development of high-pressure characterization techniques, extensive research remains to reveal the enigma surrounding this room-temperature superconductor. For instance, the isotope substitution of hydrogen with deuterium could verify its conventional superconductivity. Recently, diamagnetism associated with the Meissner effect was successfully detected in CeH$_9$ up to about 140 GPa using nitrogen-vacancy quantum sensors implanted directly within anvil[72]. Extending such measurements above 200 GPa could enable direct magnetic characterization of room-temperature superconducting LaSc$_2$H$_{24}$. In addition, advanced ultra-high-pressure tunneling spectroscopy may uncover the superconducting gap, providing insight into its superconducting nature[73]. Nevertheless, the successful experimental synthesis and characterization of LaSc$_2$H$_{24}$ are shining through the starry sky of mankind's ongoing efforts to explore high-temperature superconductors and are greatly encouraging for the exploration of wide-ranging multinary superhydrides that might exhibit higher $T_c$ even under moderate pressures.

**Conclusions**

In summary, we synthesized the room-temperature superconducting La-Sc alloy hydride under high pressure. Superconductivity with a maximum $T_c$ of 298 K was convincingly confirmed by both the zero resistance and the suppression of $T_c$ under external magnetic fields. XRD measurements indicate that the alloy hydride matches up well with the stoichiometric LaSc$_2$H$_{24}$ and remains metastable above ~194 GPa at room temperature. Our findings constitute a significant advance in condensed matter physics, demonstrating the room-temperature superconductor with a resolved crystal structure and a clearly identified superconducting mechanism, while highlighting the tremendous potential of multinary superhydrides for realizing high-temperature superconductors.

**Methods**

**Alloy precursor preparation**

The precursor La-Sc alloy was prepared either by melting or double target co-magnetron sputtering. In the melting method, La and Sc (in a molar ratio of ~1:2) with a purity of 99.9% were thoroughly heated and melted in an Ar atmosphere. To ensure compositional uniformity, the ingot was turned over and remelted multiple times. For magnetron sputtering, La and Sc metals were co-sputtered onto a glass substrate, with both targets pre-sputtered to remove surface oxides. The ~1:2 molar ratio and compositional homogeneity of La-Sc alloys were confirmed by inductively coupled plasma atomic emission spectroscopy (Thermo Fisher iCAP PRO) and scanning electron microscope (JEM-2000FS, Regulus 8100, and ZEISS Gemini 300) equipped for the energy dispersive X-ray spectroscopy (EDS). Corresponding EDS results are summarized in Extended Data Table. 1.

**Hydride synthesis under high pressure**

La-Sc alloy hydride was synthesized via the reaction of La-Sc alloy with $NH_3BH_3$ (Sigma-Aldrich, 97%) in a diamond anvil cell (DAC). The diamond anvils had culets with diameter of 30 $\mu$m, beveled at 8.5° to a diameter of about 250 $\mu$m. An insulating composite gasket was prepared following a procedure similar to that reported previously[13]. Briefly, a rhenium gasket (thickness of 250 $\mu$m) was precompressed to about 20 GPa to make an indent, from which a ~50 $\mu$m diameter cavity was drilled. The cavity was filled with an epoxy-alumina powder composite and compressed to a thickness of 10-15 $\mu$m between the anvil and rhenium gasket. Finally, a hole (diameter of 10-15 $\mu$m) was drilled to clamp the sample. A La-Sc alloy foil (~1-2 $\mu$m thickness, ~15 $\mu$m diameter) was then sandwiched between layers of $NH_3BH_3$.

Sample preparation and loading were carried out inside an Ar-filled glove box, with the residual oxygen and water levels below 0.01 ppm, ensuring the sample was properly protected from oxidation and moisture. The precursors were then compressed to the target pressure at room temperature. Pressures were determined from the room-temperature first-order diamond Raman edge calibrated by Akahama[74][75]. Two-sided laser-heating experiments were performed using a pulsed YAG laser (1064 nm) with a spot size of ~10 $\mu$m in diameter.

**Synchrotron X-ray diffraction measurements**

*In-suit* X-ray diffraction (XRD) patterns were collected at Shanghai Synchrotron Radiation Facility (SSRF) Beamlines BL15U1 ($\lambda$=0.6199 Å) and BL11ID (0.4169 Å) using focused monochromatic X-ray beams with spot sizes around 4×5 and 2×3 $\mu m^2$, respectively. Angle-dispersive XRD data were recorded with a Mar165 CCD detector (BL15U1) and a Pilatus CdTe 2M detector (BL11ID). The sample-to-detector distance and other geometric parameters were calibrated using a CeO$_2$ standard. Powder diffraction rings were integrated and converted to one-dimensional profiles using the software package Dioptas[76]. Full profile analysis and Rietveld refinements were done using GSAS-II and EXPGUI packages[77].

**Electrical transport measurements**

Resistances were measured via the four-probe van der Pauw method to eliminate contributions from sources other than the samples, with currents of 0.1-5 mA. Four hand-ranged Pt electrodes with a thickness of ~2 $\mu$m were placed in direct contact with the sample at the center of the culet. Pt foils attached to the anvil sidewall were connected to the electrodes, and enameled copper wire extended the connections to external equipment. Electronic transport measurements under external magnetic fields of cells 3, 4, and 5 were carried out using two Cryomagnetics'C-Mag Cryogen-FREE magnet systems capable of generating fields up to 12 T and 9 T. Due to a temperature gradient between the temperature sensor and the sample, the cooling and warming cycles differed[20]; onset temperatures of the resistance drops were recorded upon warming for comparisons with the other superconducting hydrides, with a slow heating ratio of 1-3 K/min and intervals of 0.1-0.5 K. Chamber pressures typically increased after the cooling cycle, the post-cooling pressure was used to represent the entire process.

**Data availability**

The authors declare that the main data supporting the findings of this study are contained within the paper and its associated Supplementary Information. All other relevant data are available from the corresponding author upon reasonable request.

**Acknowledgements**

This research was supported by the National Natural Science Foundation of China (Grant Nos. 52288102, 52090024, 12474010, 12374007, and 12474223) and Scientific Research Innovation Capability Support Project for Young Faculty (No. ZYGXQNJSKYCXNLZCXM-M12). The XRD measurements were performed at beamlines BL15U1 (31124.02.SSRF.BL15U1) and BL11ID of SSRF. Additional support was provided by beamline BL10XU of SPring-8 and the User Experiment Assist System of SSRF (31124.02.SSRF.LAB).

**Author contributions**

Yanming Ma and Guangtao Liu designed the research; Yinggang Song and Chuanheng Ma prepared the samples and conducted the electrical measurements with the assistance from Guangtao Liu, Hongbo Wang, Mi Zhou, and Yanpeng Qi. Yinggang Song, Chuanheng Ma, and Guangtao Liu performed X-ray diffraction and processed the corresponding data with the assistance from Hongbo Wang, Mi Zhou, and Shourui Li. Guangtao Liu, Yinggang Song, and Chuanheng Ma analyzed the experimental results. Hanyu Liu performed the theoretical analysis. Yinggang Song, Guangtao Liu, and Yanming Ma wrote the manuscript with input from all co-authors. All of the authors, along with other related work, were carried out under the meticulous guidance of Yanming Ma.

**Competing interests**

The authors declare no competing interests.

These authors contributed equally: Yinggang Song, Chuanheng Ma, and Hongbo Wang.

e-mail: liuguangtao@jlu.edu.cn; mym@jlu.edu.cn

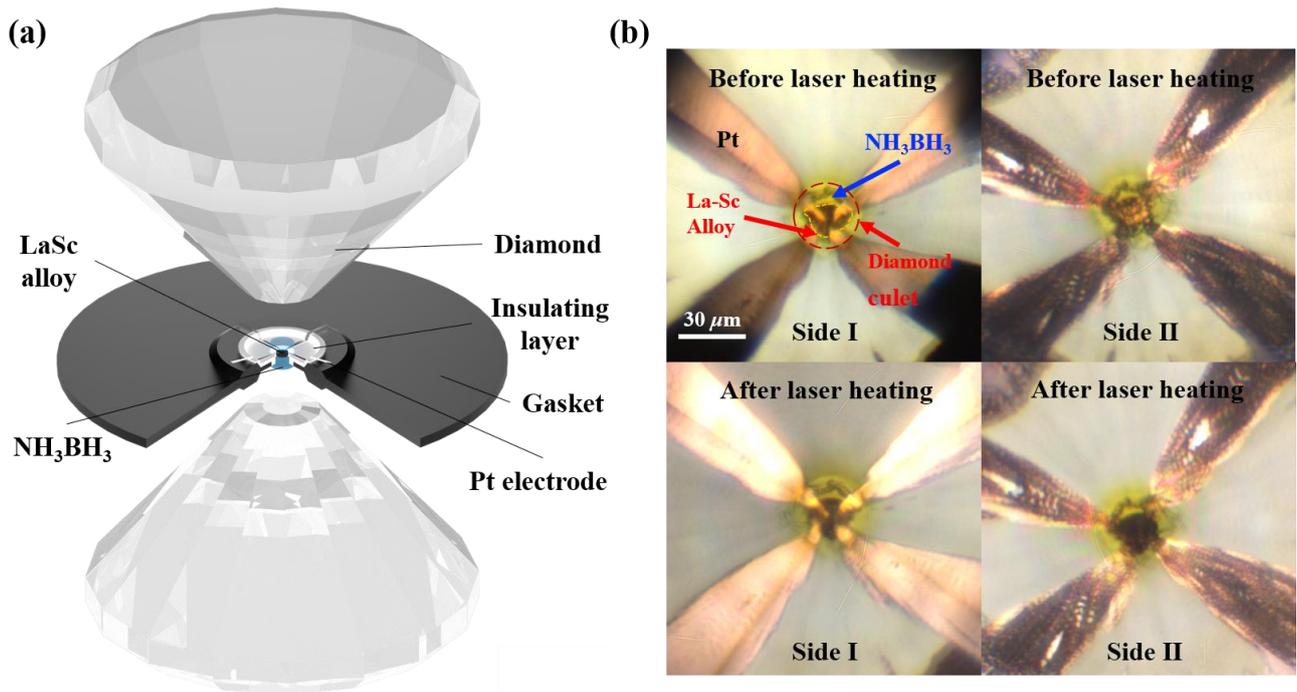

Fig. 1: (a) Schematic of diamond anvil and sample loading details. In diamond anvil cell, the La-Sc alloy is sandwiched between the hydrogen source $NH_3BH_3$ and attached to the electrodes on insulating layer. (b) Optical micrographs of the sample chambers containing $NH_3BH_3$ and Pt electrodes in cell 4 before and after double-sided laser heating. The top and bottom are photos of cell 4 before and after laser heating from two sides, respectively. The edges of the La-Sc alloy precursor and diamond culet are marked with yellow and red dotted lines, respectively. The blue arrow points to the transparent $NH_3BH_3$ in sample chamber.

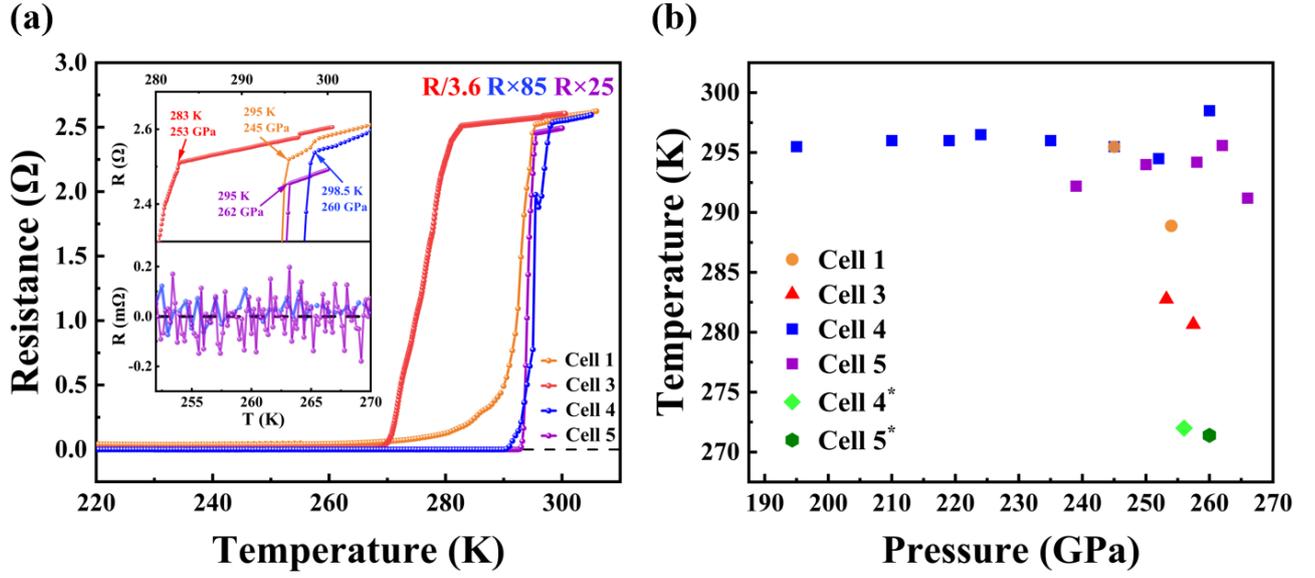

Fig. 2: (a) Electrical transport measurements of the synthesized superhydride in cells 1, 3, 4, and 5 under different pressures. The upper-left insert shows the enlarged transition area, and the $T_c^{onset}$s in independent cells are marked next to the data. The resistance data with near-zero values in cells 4 and 5 are shown on a smaller scale in the lower-left inset. (b) The dependence of $T_c^{onset}$ on pressure from different cells. Different colors distinguish data from cell 1, cell 3, cell 4, and cell 5, respectively. $T_c^{onset}$ is defined as the temperature at which the resistance begins to drop. The $T_c$s in cells 4* and 5* correspond to those observed in the first measurements after synthesis.

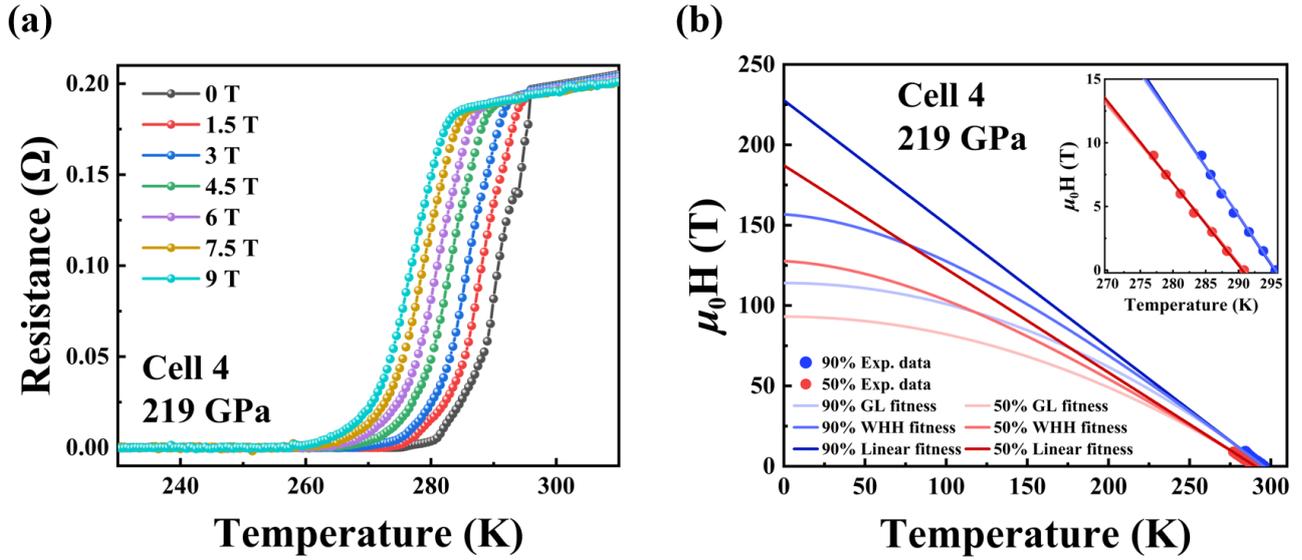

Fig. 3: (a) Raw data of temperature dependence of the electrical resistance in cell 4 under applied magnetic fields of H=0, 1.5, 3, 4.5, 6, 7.5, and 9 T at 219 GPa. (b) Upper critical fields as a function of temperature following the criteria of 90% (blue markers) and 50% (red markers) of the resistance in the metallic state at 219 GPa, fitted with the GL, WHH models, and linear fitness. The insert enlarges the fitness of experimental data.

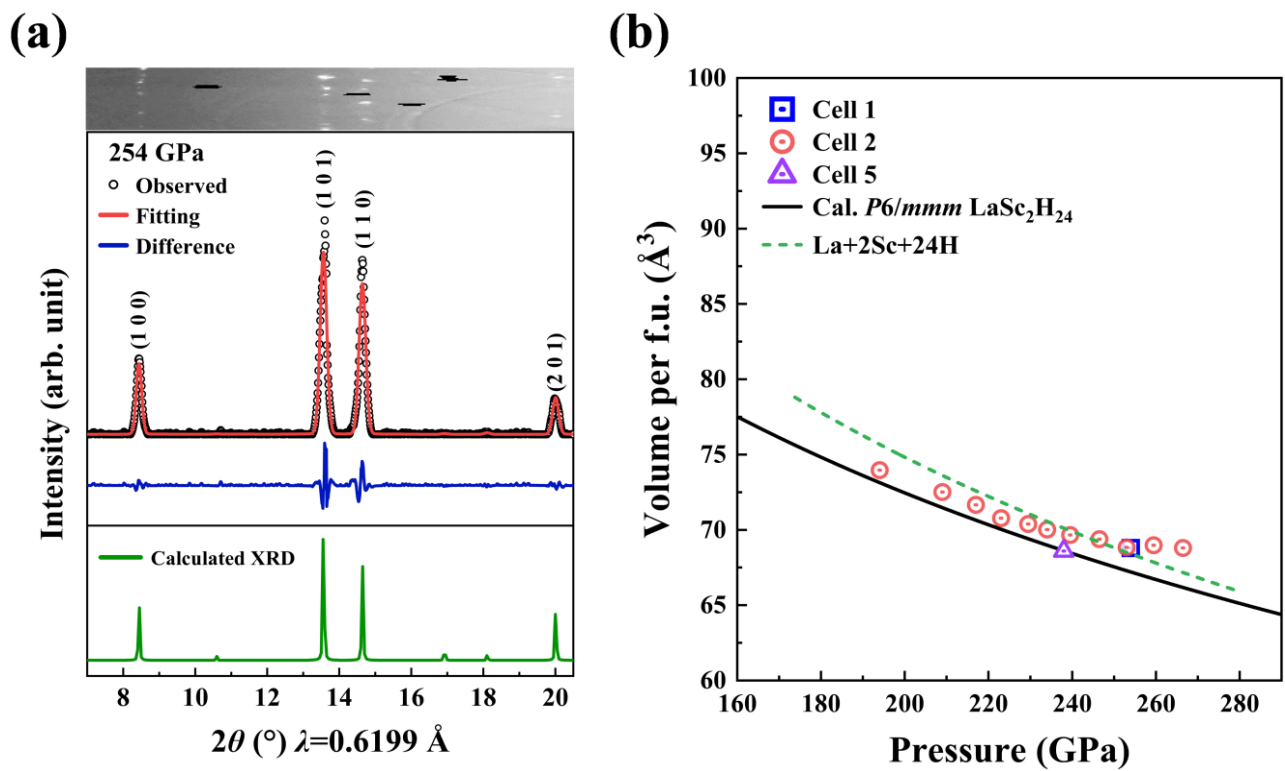

Fig. 4: (a) Synchrotron x-ray diffraction pattern of the heated sample in cell 1 at 254 GPa and the Rietveld refinement of the *P6/mmm* LaSc$_2$H$_{24}$ structure. The overexposed points caused by single-crystal-like diffraction are masked as black areas in the raw XRD pattern on the top of (a), and the continuous background is removed before performing the integration and Rietveld analysis. The blue square and red and blue curves correspond to the experimental data, Rietveld refinement fit, and residue, respectively. The green curve indicates the calculated peak positions and calculated XRD, which shows the relative intensity of each peak for *P6/mmm* LaSc$_2$H$_{24}$. The crystal surface indexes are marked next to the peaks. (b) Experimental equation-of-state data for the synthetic samples are compared with the theoretical data (solid black line) derived from the theoretical EOS of *P6/mmm* structured LaSc$_2$H$_{24}$ (Ref. [1]). The experimental data from cells 1, 2, and 5 are marked with black, red, and purple symbols, respectively. The green dashed curve shows that for the unreacted elemental assemblage of La, 2Sc, and 24H. The ~5 GPa error in pressure comes from the pressure gradient on the diamond culets and the Raman edge reading in pressure determination. The volume errors are displayed in Extended Data Table. 3.

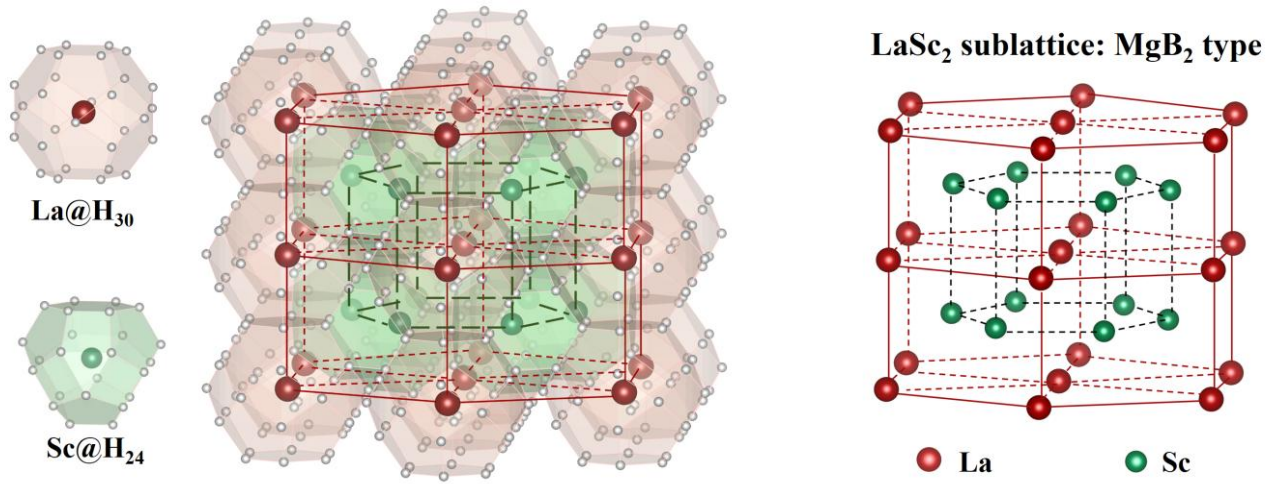

Fig. 5: The crystal structure of $P6/mmm$ LaSc$_2$H$_{24}$ consisting of H$_{30}$ cages and H$_{24}$ cages. Red, green, and silver balls represent La, Sc, and H atoms, respectively. Red La atom and green Sc atom are surrounded by H$_{30}$ cage and H$_{24}$ cage, respectively.

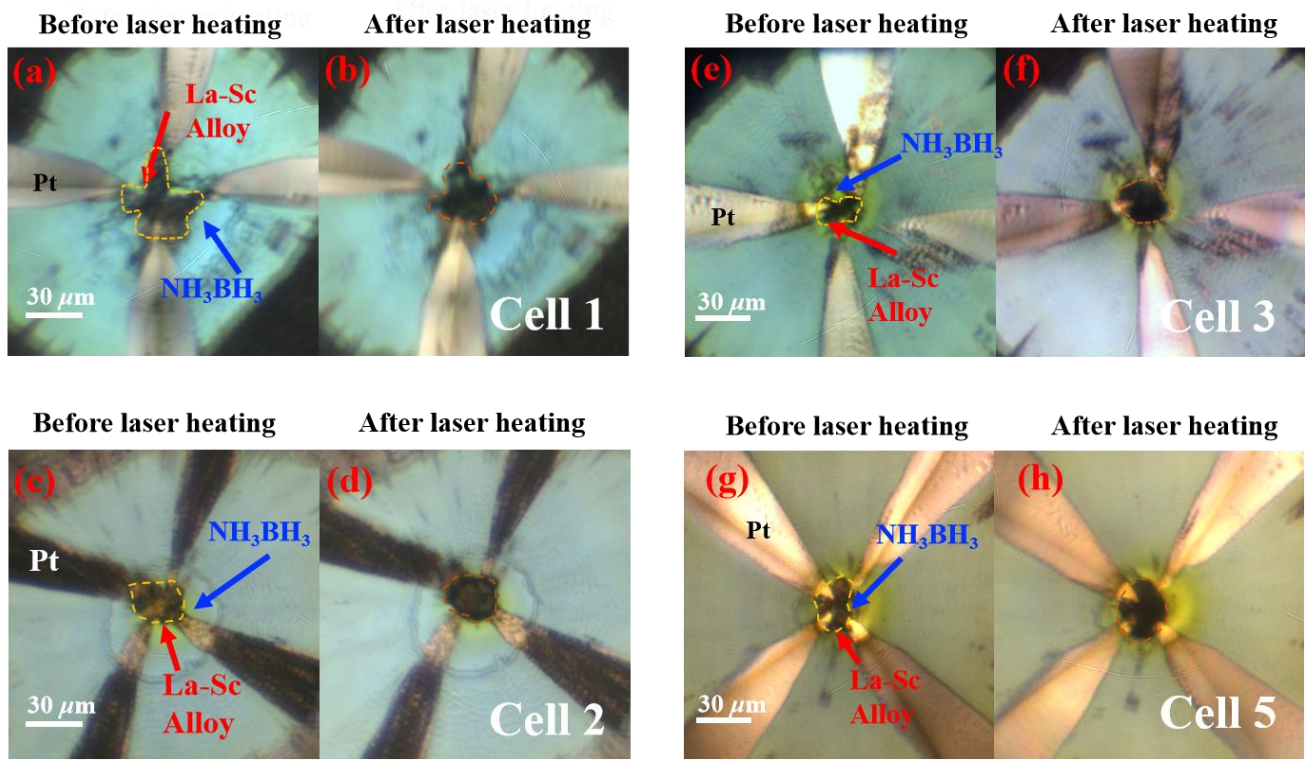

**Extended Data Fig. 1** Optical micrographs of the sample chambers containing NH$_3$BH$_3$ and Pt electrodes in cells 1, 2, 3, and 5 before and after laser heating. (a), (c), (e), and (g) are photos before laser heating, respectively. (b), (d), (f), and (h) are photos after laser heating, respectively. The edges of La-Sc alloy precursor and product are marked with yellow dotted lines. The blue arrow points to the transparent NH$_3$BH$_3$ in sample chamber.

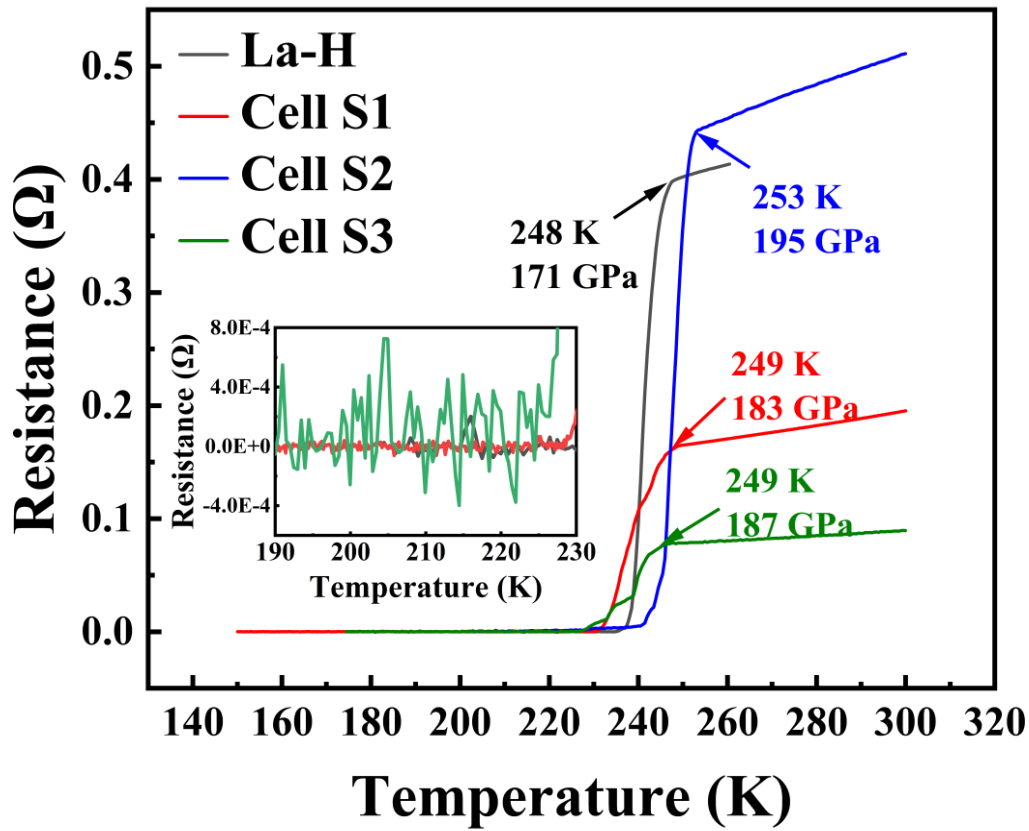

**Extended Data Fig. 2.** Raw data of temperature dependence of resistance of the sample for the warming cycles from cells La-H, S1, S2, and S3. The $T_c$s of 248, 249, 253, and 249 K were observed in cells S1, S2, and S3, respectively. The resistance data with near-zero values in cells La-H, S1, and S3 are shown on a smaller scale in the inset.

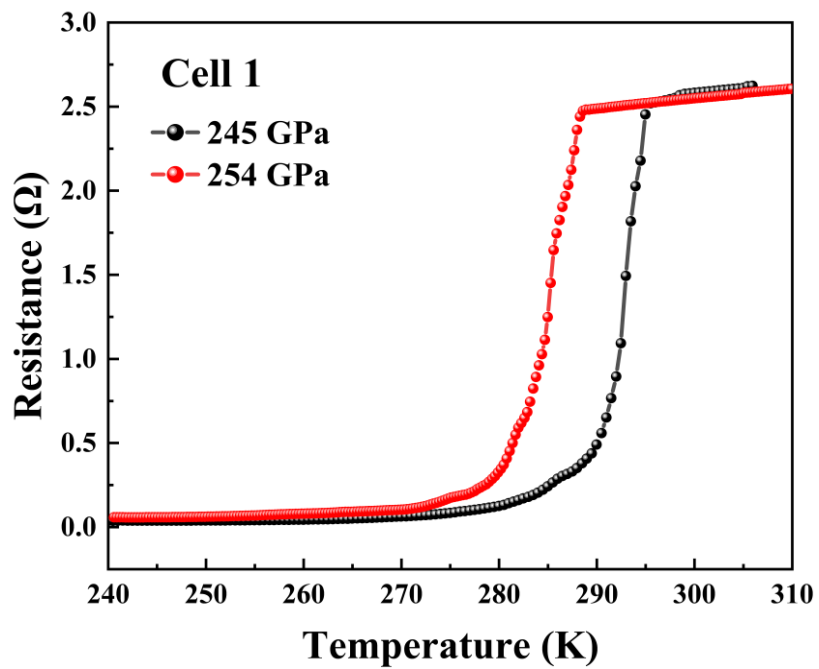

**Extended Data Fig. 3.** Raw data of temperature dependence of resistance of the sample for the warming cycles from cell 1. The black and red curves represent the resistance data with the temperature dependence at 245 and 254 GPa, respectively, during the compression.

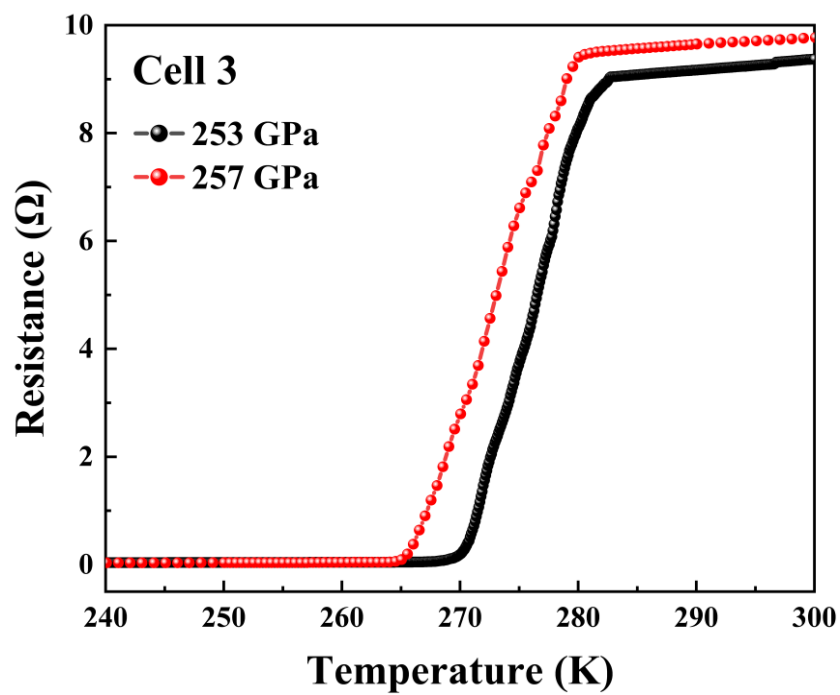

**Extended Data Fig. 4** Raw data of temperature dependence of resistance of the sample for the warming cycles from cell 3. The black and red curves represent the resistance data with the temperature dependence at 253 and 257 GPa, respectively, during the compression.

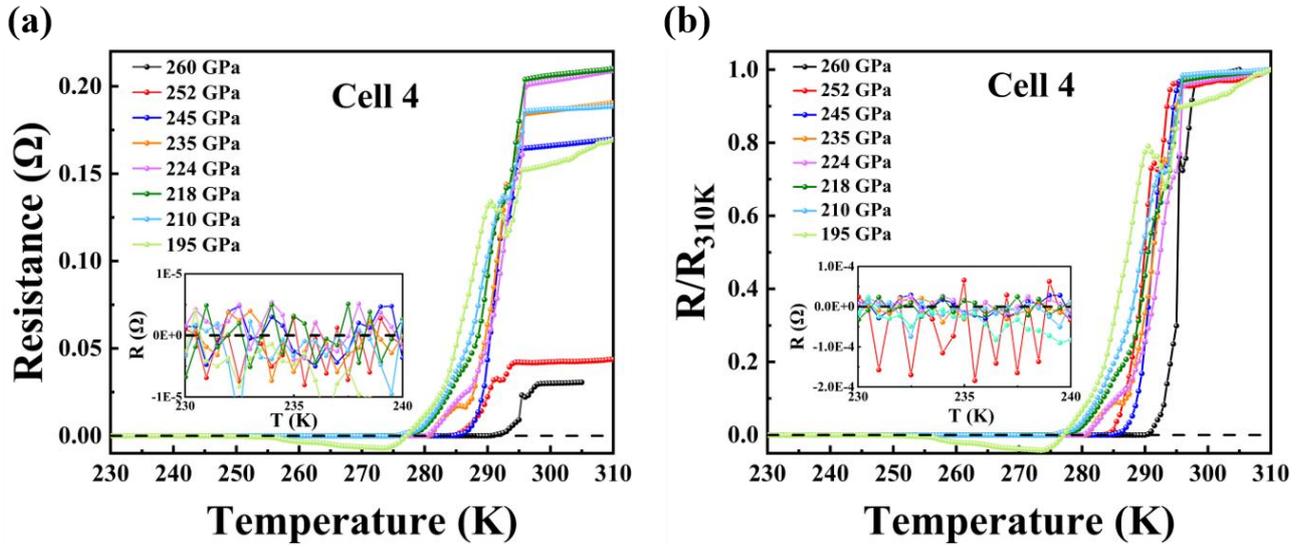

**Extended Data Fig. 5** (a) Raw data of temperature dependence of resistance of the sample for the warming cycles at different pressures from cell 4. The resistance data with near-zero values are shown on a smaller scale in the left inset. (b) The vertical axis is the resistance divided by the resistance at 310 K for comparison. The resistance data with near-zero values are shown on a smaller scale in the left inset.

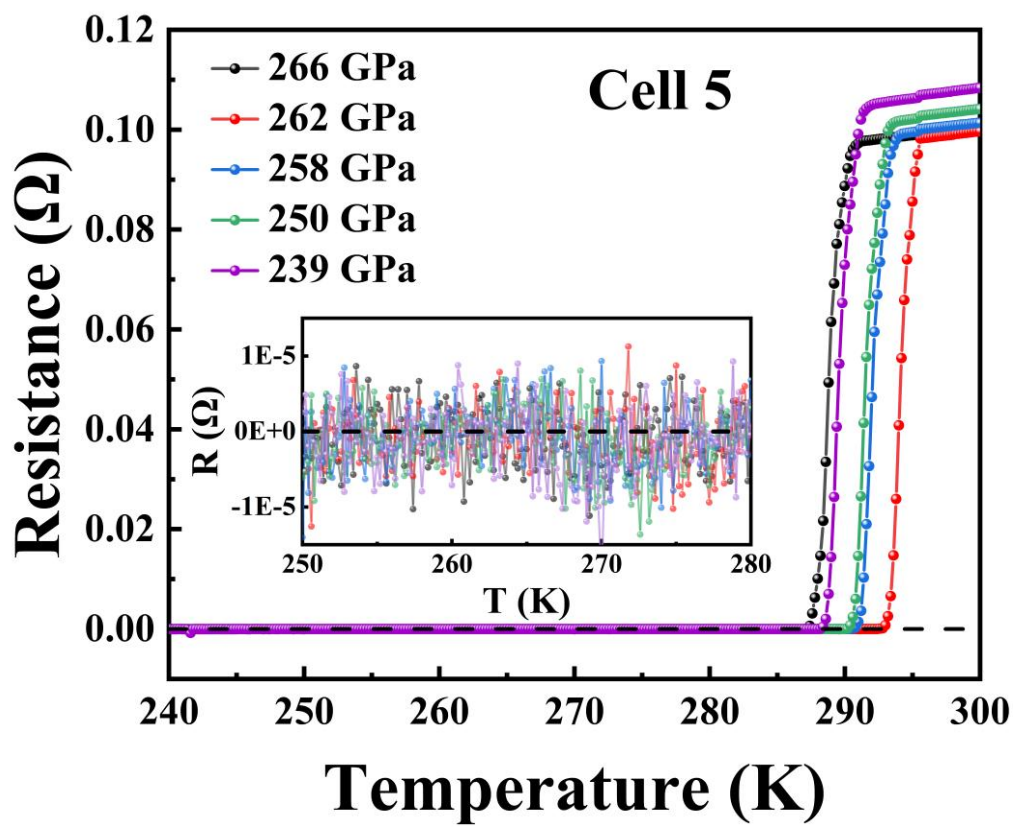

**Extended Data Fig. 6** Raw data of temperature dependence of resistance of the sample for the warming cycles at different pressures from cell 5. The resistance data with near-zero values are shown on a smaller scale in the inset.

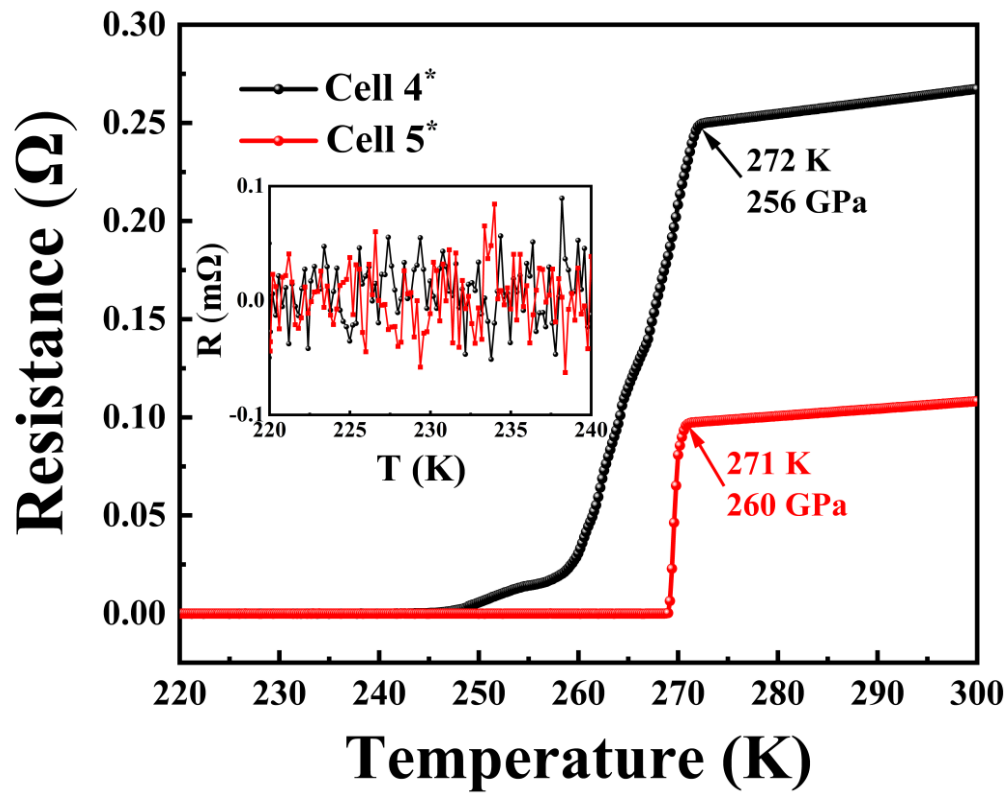

**Extended Data Fig. 7.** Raw data of temperature dependence of resistance of the sample for warming cycle from cells 4* and 5* at 256 and 260 GPa, respectively. The resistance data with near-zero values are shown on a smaller scale in the inset.

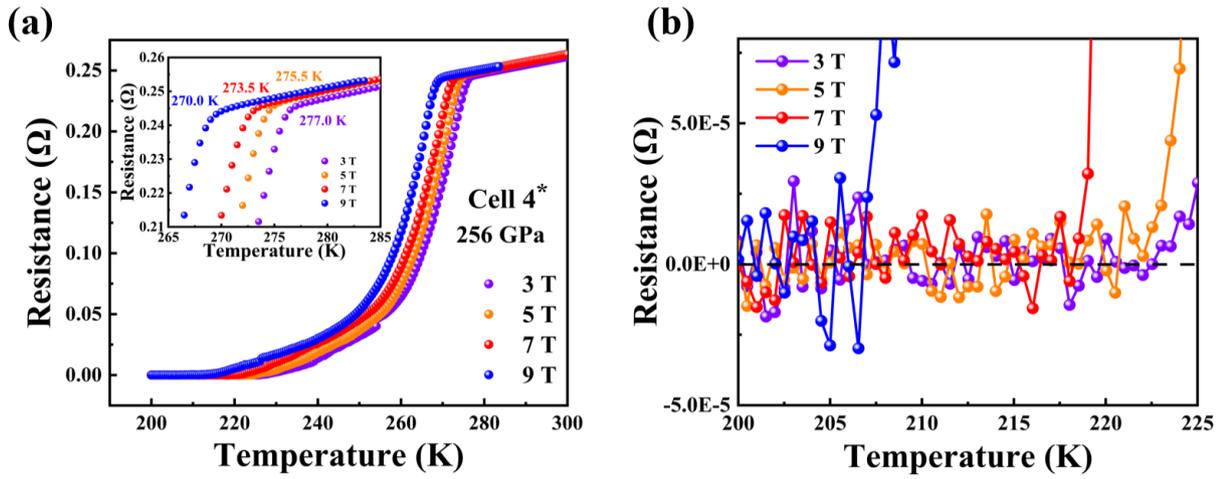

**Extended Data Fig. 8** (a) Raw data of temperature dependence of the electrical resistance in cell 4* under applied magnetic fields of H=3, 5, 7, and 9 T at 256 GPa. The insert shows the enlarged transition area, and the superconducting transition temperatures at different applied magnetic fields are marked next to the data. (b) Raw data of resistance data with near-zero values under different applied magnetic fields.

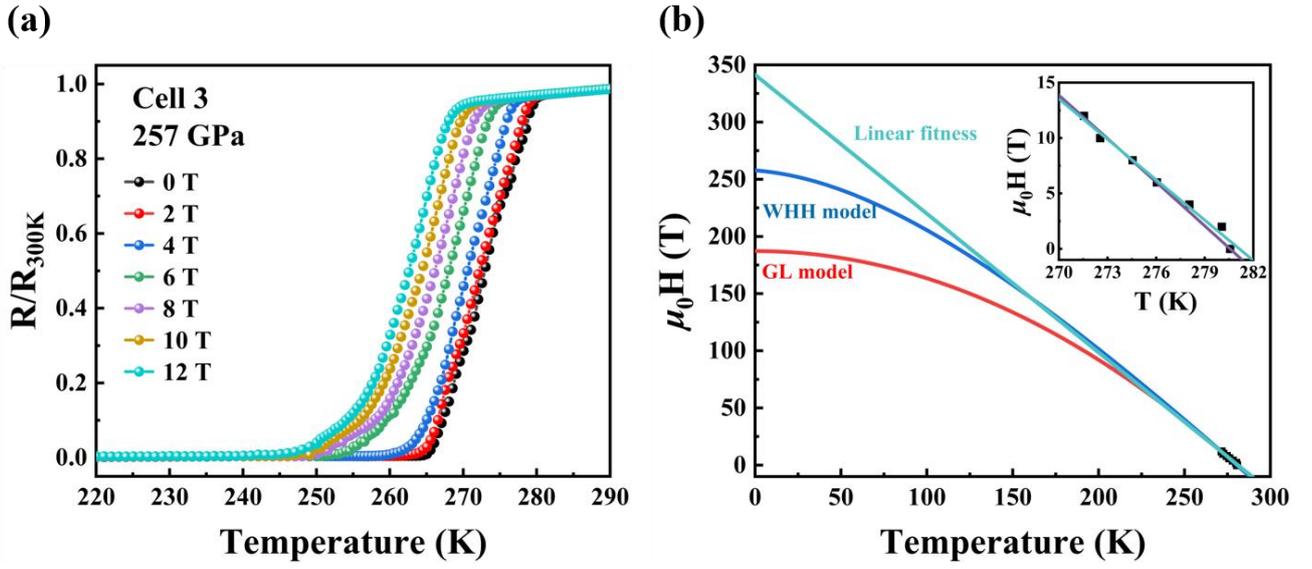

**Extended Data Fig. 9**: (a) Raw data of temperature dependence of the electrical resistance in cell 3 under applied magnetic fields of H=0, 2, 4, 6, 8, 10, and 12 T at 257 GPa. The vertical axis is the resistance divided by the resistance at 300 K. Upon the increasing magnetic fields, the $T_c$ gradually decreased about 9 K from 281 K at 0 T to 272 K at 12 T. (b) The upper critical fields are $\mu_0 H_{c2}(T)$ towards 0 K are 187 T, 258 T, and 341 T, using GL, WHH, and linear fitness, respectively. The Ginzburg-Landau coherence length deduced from the $H_{c2}(0)$ is 1.33 nm.

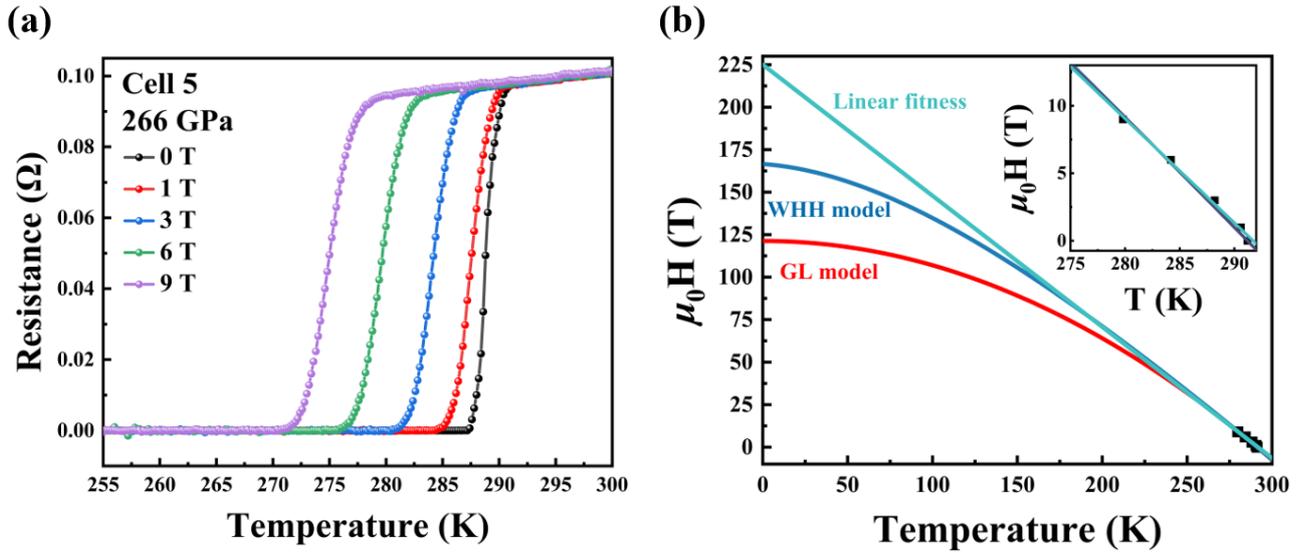

**Extended Data Fig. 10**: (a) Raw data of temperature dependence of the electrical resistance in cell 5 under applied magnetic fields of H=0, 1, 3, 6, and 9 T at 266 GPa. Upon the increasing magnetic fields, the $T_c$ gradually decreased about 11 K from 291 K at 0 T to 280 K at 9 T. (b) The upper critical fields are $\mu_0 H_{c2}(T)$ towards 0 K are 121 T, 166 T, and 225 T, using GL, WHH, and linear fitness, respectively. The Ginzburg-Landau coherence length deduced from the $H_{c2}(0)$ is 1.65 nm.

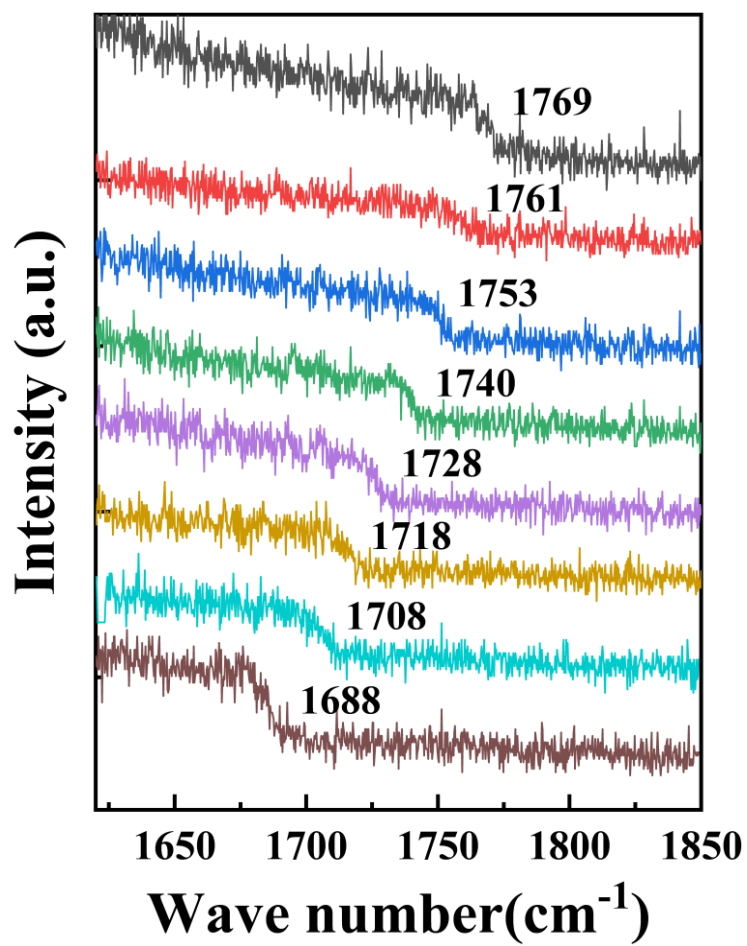

**Extended Data Fig. 11** Detailed diamond Raman spectra at different pressures during decompression in cell 4. The numbers represent the wave number read at 1/3 of the edge at different pressures.

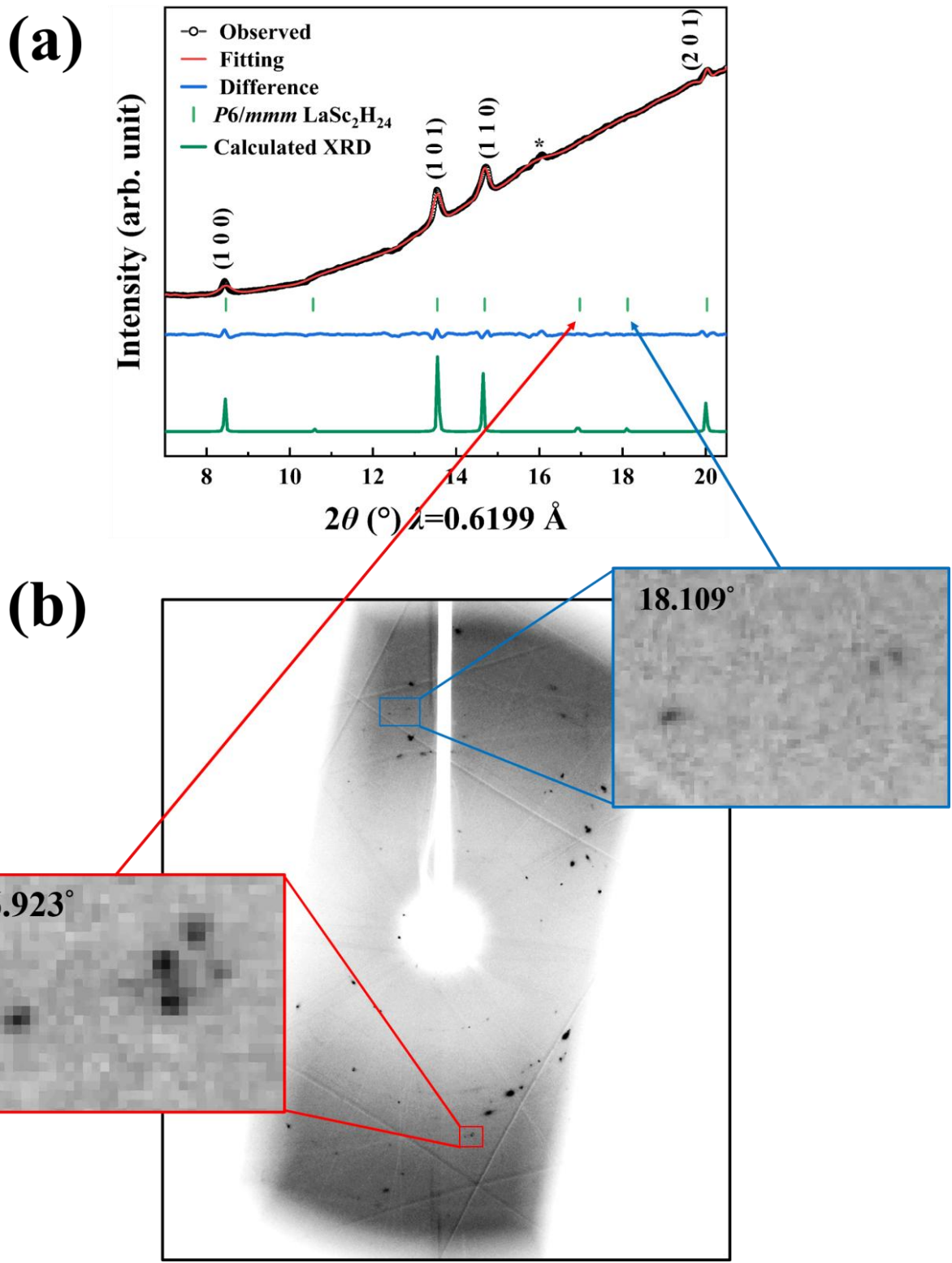

**Extended Data Fig. 12** (a) Synchrotron x-ray diffraction pattern of the heated sample in cell 1 at 254 GPa and the Rietveld refinement of the $P6/mmm$ $LaSc_2H_{24}$ structure. The green curve and sticks represent the simulated XRD and diffraction peaks of $P6/mmm$ $LaSc_2H_{24}$. (b) The raw XRD pattern obtained from cell 1. Red and Blue squares enlarge the areas where the spots are observed at 16.923° and 18.109°, respectively.

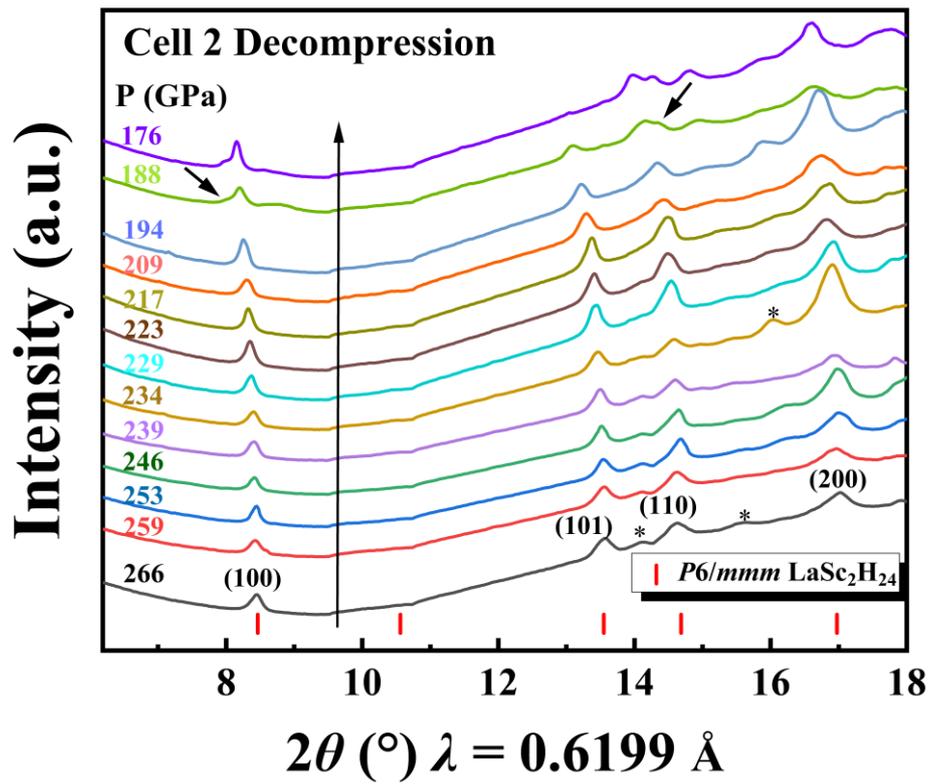

**Extended Data Fig. 13** Synchrotron X-ray diffraction patterns in the pressure range of 266-176 GPa (cell 2). The red vertical lines mark the diffraction peak positions from $P6/mmm$ LaSc$_2$H$_{24}$. The Miller indexes are marked next to the peaks. The weak peaks marked with asterisks may be from other undetermined hydride(s). Black arrows point to the split of diffraction peaks (1 0 0) and (1 1 0) at 188 GPa.

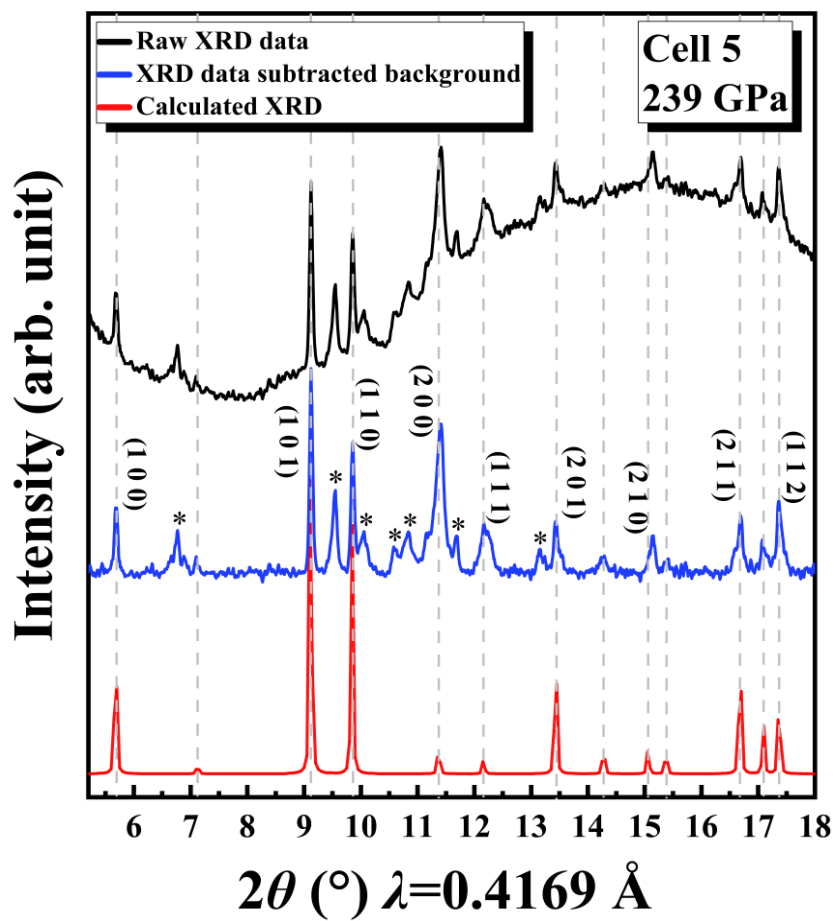

**Extended Data Fig. 14** Synchrotron XRD pattern in cell 5 at 239 GPa. The black, blue, and red curves represent the raw XRD data, XRD data after background subtraction, and calculated XRD of the $P6/mmm$ $LaSc_2H_{24}$ structure. The grey vertical dashed lines represent diffraction peaks of $P6/mmm$ $LaSc_2H_{24}$. The crystal surface indexes are marked next to the peaks. The peaks marked with asterisks may be from other undetermined hydride(s).

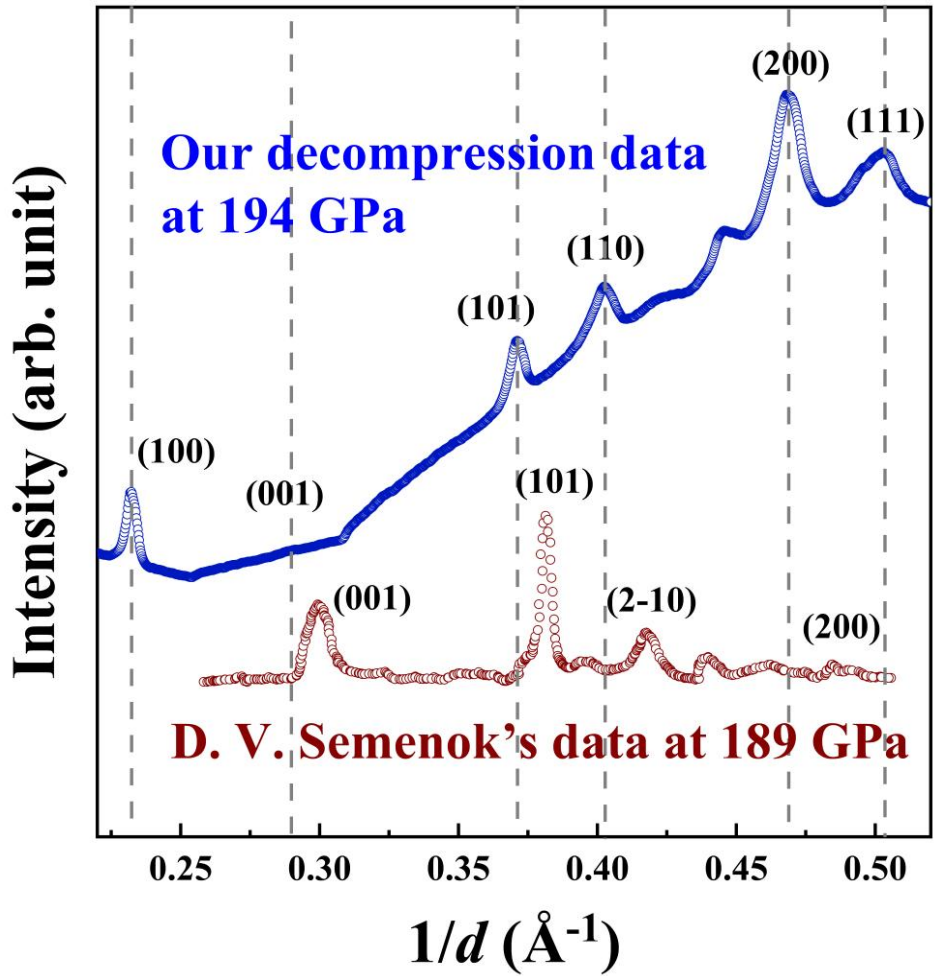

**Extended Data Fig. 15:** Blue points represent our decompression data at 194 GPa, and red points represent the data of Semenok's at 189 GPa. Vertical gray dashed lines indicate the diffraction peaks of our *P*6/*mmm* LaSc$_2$H$_{24}$ phase, with corresponding peaks from our data. Diffraction angles (2θ) are converted to 1/d to enable comparison across different X-ray wavelengths.

# Supplementary Tables

**Extended Data Table 1.** Scanning electron microscope (SEM) images, the energy-dispersive x-ray spectrum (EDX), and map results of La-Sc alloy.

| Alloy | SEM image | La | Sc | EDX | Map results (La: Sc atom%) |
|---|---|---|---|---|---|
| #1 (Melted method) | 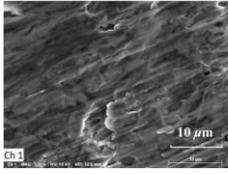 | 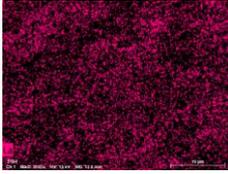 | 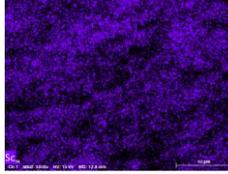 | 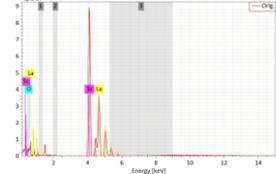 | 31.03: 68.97 |
| #2 (Magnetron sputtering) | 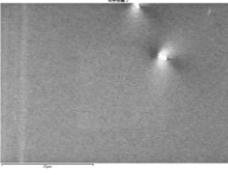 | 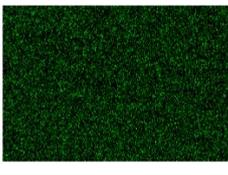 | 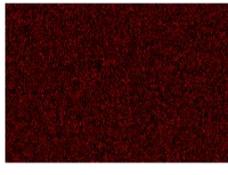 | 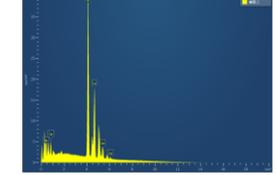 | 34.36: 65.64 |
| #3 (Magnetron sputtering) | 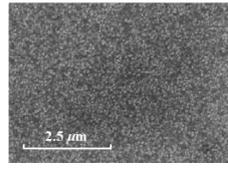 | 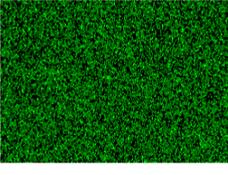 | 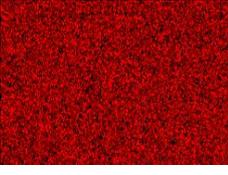 | 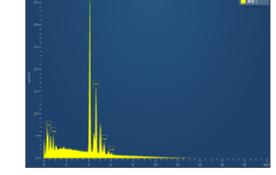 | 34.14: 65.86 |

**Extended Data Table 2.** Details of the five DACs and 13 measurement runs.

| Number | Culet size ($\mu$m) | Gasket | Precursor | Pressure (GPa) | Measurement run |
|---|---|---|---|---|---|
| #Cell-1 | 30 | Re+Al$_2$O$_3$+epoxy | La-Sc alloy +NH$_3$BH$_3$ | 245-254 | R-T*, XRD |
| #Cell-2 | 30 | Re+Al$_2$O$_3$+epoxy | La-Sc alloy +NH$_3$BH$_3$ | 253-257 | XRD |
| #Cell-3 | 30 | Re+Al$_2$O$_3$+epoxy | La-Sc alloy +NH$_3$BH$_3$ | 176-266 | R-T, MF** |
| #Cell-4 | 30 | Re+Al$_2$O$_3$+epoxy | La-Sc alloy +NH$_3$BH$_3$ | 260 | R-T, MF |
| #Cell-4* | 30 | Re+Al$_2$O$_3$+epoxy | La-Sc alloy +NH$_3$BH$_3$ | 256 | R-T, MF |
| #Cell-5 | 30 | Re+Al$_2$O$_3$+epoxy | La-Sc alloy +NH$_3$BH$_3$ | 239-266 | R-T, MF, XRD |
| #Cell-5* | 30 | Re+Al$_2$O$_3$+epoxy | La-Sc alloy +NH$_3$BH$_3$ | 260 | R-T |

Cell-4* and Cell-5* represent the samples that underwent initial laser heating, where $T_c$s of 272 K and 271 K were observed.

Upon re-laser heating and measurement, Cell-4 (the same cell as Cell-4*) yielded a higher $T_c$ of 296 K.

* Temperature-dependent resistance measurement (R-T).

** Transport measurements under varying external magnetic fields (MF).

**Extended Data Table 3.** Experimental cell parameters and volumes per formula of $P6/mmm$ LaSc$_2$H$_{24+x}$ at different pressures.

| Cell Number | Pressure (GPa) | $a$ (Å) | $c$ (Å) | $V$ (Å$^3$/formula) | x in LaSc$_2$H$_{24+x}$ |
|---|---|---|---|---|---|
| #Cell-1 | 254 | 4.86(4) | 3.35(6) | 68.78(8) | 0.3 |
| | 266 | 4.85(9) | 3.36(5) | 68.80(3) | 1.4 |
| | 259 | 4.87(0) | 3.35(8) | 68.97(7) | 0.9 |
| | 253 | 4.86(2) | 3.36(1) | 68.83(9) | 0.5 |
| | 246 | 4.87(7) | 3.36(8) | 69.38(8) | 0.3 |
| | 239 | 4.88(2) | 3.37(3) | 69.65(6) | 0 |
| #Cell-2 | 234 | 4.88(5) | 3.38(7) | 70.01(7) | -0.3 |
| | 229 | 4.89(2) | 3.39(5) | 70.39(1) | -0.4 |
| | 223 | 4.90(1) | 3.40(1) | 70.78(1) | -0.6 |
| | 217 | 4.93(0) | 3.40(3) | 71.66(1) | -0.6 |
| | 209 | 4.94(0) | 3.43(0) | 72.51(5) | -0.7 |
| | 194 | 4.97(1) | 3.45(5) | 73.96(6) | -1.1 |
| #Cell-5 | 239 | 4.85(9) | 3.35(5) | 68.59(9) | -0.6 |

The experimental hydrogen content is estimated by comparing the measured volumes of the corresponding high-pressure phases of elemental La (distorted-$fcc$ phase), Sc (IV and V phases), and H$_2$ (I, III, and IV phases), with x in LaSc$_2$H$_{24+x}$ subsequently determined using the following formula: $x = \frac{V - V_{La} - 2V_{Sc}}{V_H} - 24$.